\documentclass[pmlr,twocolumn,10pt]{jmlr} % useful when printing your own Letter copy; should not impact layout; do not use for submission
\usepackage{isipta2019}
\usepackage[american]{babel}
\usepackage{booktabs} % commands to create good-looking tables
\usepackage{tikz} % nice language for creating drawings and diagrams
\newcommand{\indep}{\perp \!\!\!\!\! \perp}

\jmlrworkshop{ISIPTA 2019}

\title[Vacuous Orientation Simultaneous Inference]{Simultaneous Inference Under the Vacuous Orientation Assumption}

\author{% Authors are listed one per line and grouped per affiliation
  \Name{Ruobin {Gong}}\Email{ruobin.gong@rutgers.edu}\\
  \addr Department of Statistics, Rutgers University, New Jersey, USA
}

\begin{document}
\maketitle

% --- customized commands --- %

\newcommand{\hA}{{\bf A}}
\newcommand{\hC}{{\bf C}}
\newcommand{\hE}{{\bf E}}
\newcommand{\hM}{{\bf M}}
\newcommand{\hU}{{\bf U}}
\newcommand{\hX}{{\bf X}}
\newcommand{\hY}{{\bf Y}}

\newcommand{\ha}{{\bf a}}
\newcommand{\he}{{\bf e}}
\newcommand{\hm}{{\bf m}}
\newcommand{\hu}{{\bf u}}
\newcommand{\hx}{{\bf x}}
\newcommand{\hy}{{\bf y}}
\newcommand{\hbeta}{\boldsymbol{\beta}}
\newcommand{\hepsilon}{\boldsymbol{\epsilon}}

\newcommand{\Bb}{\mathcal{B}}
\newcommand{\Dd}{\mathcal{D}}
\newcommand{\Uu}{\mathcal{U}}

\newcommand{\FF}{\mathsf{F}}
\newcommand{\RR}{\mathsf{R}}

% (p, q, r)
\newcommand{\tp}{\mathsf{p}}
\newcommand{\tq}{\mathsf{q}}
\newcommand{\tr}{\mathsf{r}}

\newcommand{\EEE}{\mathbb{E}}
\newcommand{\III}{\mathbb{I}}
\newcommand{\RRR}{\mathbb{R}}

\newcommand{\eqdef}{\;\overset{\text{def}}{=\joinrel=}\;}

\newcommand{\lp}{\underline{P}}
\newcommand{\up}{\overline{P}}
\newcommand{\RRM}[1]{\RR_{\hM \mid \EEE{#1}}}
\newcommand{\tx}{t_\hy}

\newcommand{\red}[1]{{\color{red!60!black}#1}}

% --------------------------- %

\begin{abstract}
I propose a novel approach to simultaneous inference that alleviates the need to specify a correlational structure among marginal errors. The {\it vacuous orientation} assumption retains what the normal i.i.d. assumption implies about the distribution of error configuration, but relaxes the implication that the error orientation is isotropic. When a large number of highly dependent hypotheses are tested simultaneously, the proposed model produces calibrated posterior inference by leveraging the logical relationship among them. This stands in contrast to the conservative performance of the Bonferroni correction, even if neither approaches makes assumptions about error dependence. The proposed model employs the Dempster-Shafer Extended Calculus of Probability, and delivers posterior inference in the form of stochastic three-valued logic. 
\end{abstract}
\begin{keywords}
Dempster-Shafer theory; belief function; Bonferroni correction; familywise error rate; calibrated inference
\end{keywords}

\section{Introduction}\label{sec:intro}

In scientific explorations, the analyst often needs to verify more than one hypothesis based on data from a single experiment. The multiplicity of hypotheses posits a threat to the trustworthiness of the overall conclusion. The inferential procedure employed, even if statistically valid for verifying single hypotheses, may no longer retain validity if not carefully compounded across hypotheses. The correct compounding of statistical procedures relies on adequate knowledge about the dependence relationship between the observed data as well as the hypotheses, neither of which is likely available to the analyst.

For concreteness, let $\hM = \left(M_1, \ldots, M_k \right)$ be a vector of unknown parameters, and $\hY = \left(Y_1,\ldots, Y_k\right)$ a vector of observable data that aims to measure $\hM$. Suppose for $i = 1,\ldots, k$,
\begin{equation}\label{eq:sampling}
Y_i \sim N \left(M_i, \, S^2\right).	
\end{equation}
That is, each $M_i$ is measured exactly once by $Y_i$, with a normally distributed measurement error of variance $S^2$. Suppose for now that $S^2$ is known. The goal is to make inference about the uncertain values of $\hM$ without additional prior information.

It seems intuitive that for an individual $M_i$, a best guess at its value is $Y_i$, accompanied by a confidence statement as an interval centered at $Y_i$ with width proportional to $S$. This follows from the sampling model in (\ref{eq:sampling}). However, to make confidence statements about $\hM$, (\ref{eq:sampling}) alone is not enough. Missing from the specification is a dependence structure among the observations given their unknown true means. 

\begin{figure}
\floatconts{fig:iid}
{\caption{The normal i.i.d. sampling model of $\hY$ given $\hM$ when $k = 2$. The i.i.d. assumption implies that 1) the sum of squares (configuration) of measurement errors is distributed $\chi_2^2$, and 2) their orientations are isotropic, i.e. uniform in all directions.}}
{\label{fig:iid}
\includegraphics[width=\linewidth]{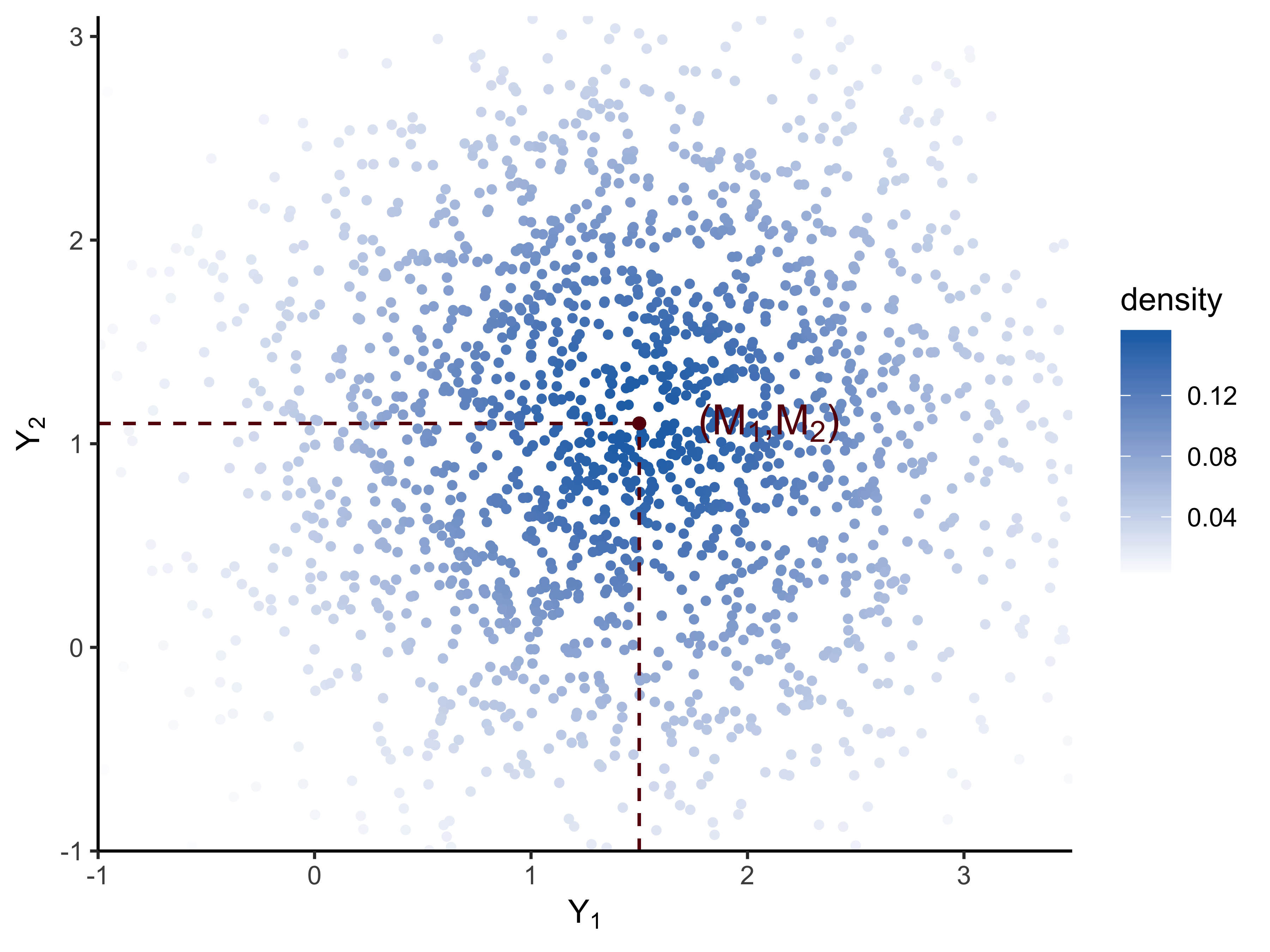}
}
\end{figure}

The scenario described here is an abstraction of a typical scientific experiment. A total of $k$ unknown quantities are learned at the same time, and each $Y_i$ is a summary statistic. As part of the inferential procedure, the marginal reference distributions of the $Y_i$'s are usually well-understood. However, information regarding their interdependence is much harder to come by. Often out of convenience, or perhaps a lack of better alternative, it is assumed that
\begin{equation}\label{eq:indep}
	Y_i \indep Y_j \mid \hM \quad \forall i \neq j,
\end{equation}
hence $\hY$ is multivariate normal with mean $\hM$ and covariance proportional to the identity matrix. We refer to (\ref{eq:sampling}) and (\ref{eq:indep}) as the sampling model under the {\it normal i.i.d.} (or just i.i.d.) assumption. Figure~\ref{fig:iid} is an illustration of it for $k=2$. 

For problems of higher dimensions, computation can be vastly simplified if some form of independence can be assumed. Within the context of multiple hypothesis testing, a most widely adopted procedure to control the {false discovery rate} is the Benjamini-Hochberg procedure \cite{benjamini1995controlling}, which is valid under the assumption of independence or mild positive dependence \cite{benjamini2001control}. However, independence (or for that matter, any known structure of dependence) among hypotheses is all but likely to hold. In particular, independence is easily violated when the parameters of interest and the summary statistics are devised sequentially according to previous observations and verification results. Such a trajectory is nevertheless a typical one in scientific explorations.

Things become more complicated when an even larger collection of hypotheses involving $\hM$, highly collinear among themselves, are to be verified together. For example, one can test whether all pairwise contrasts are equal to zero: $M_i = M_j$, which makes a total of $k(k-1)/2$ comparisons involving only $k$ unknown quantities. This case is investigated in Example~\ref{ex:contrast} in Section~\ref{sec:posterior}. Once the data is observed, one is always guaranteed to find {\it some} function $g$ about the parameter, however obscure and scientifically insignificant it may be, such that the observed data produces strikingly strong support for the statement $g\left(\hM\right)=0$. This illegitimate maneuver is called ``data-snooping'', and it invalidates the nominal significance level claimed by the test. Post-hoc analyses aim to avoid it by controlling for the {\it familywise error rate}, that is, the probability of making at most one false rejection among a family of hypothesis tests when the nulls are all true. 

A classic procedure to control for the familywise error rate is the Bonferroni correction \cite{dunn1961multiple}. If $p$ hypotheses are tested together, the required significance level for each individual test is reduced to $\alpha/p$, such that the familywise error rate is no more than $\alpha$. Notably, the validity of the Bonferroni correction does not rely on an independence assumption about the hypotheses. However, it does not take into consideration the logical structure behind the hypotheses, contributing to its well-known conservative behavior. More discussions and comparisons involving the Bonferroni procedure can be found in Section~\ref{subsec:rectangle}.

From an estimation point of view, the model in (\ref{eq:sampling}) poses challenges to both the likelihood and Bayesian approaches. Since each $M_i$ is measured exactly once, no large sample asymptotic approximation to the likelihood is appropriate, even when the variance parameter $S^2$ is known. In order to obtain a distributional description of $\hM$, one must resort to Bayesian methods and assume a prior for the parameter vector $\hM$. Again due to the small sample size, any prior will exude significant influence over the posterior for $\hM$, which is undesirable unless a certain shrinkage effect is intentionally sought for. If $S^2$ is not known {\it a priori}, the maximum likelihood estimate for $S^2$ does not exist, whereas Bayesian methods need a prior for $S^2$ that cannot be updated based on the data. Challenges regarding unknown $S^2$ is addressed in Section~\ref{sec:unknown-s}. Lastly, if the dependence structure of $\hY$ is not precisely available, both likelihood and Bayesian methods will fail to prescribe a variance quantification in their respective inference for $\hM$.

\paragraph{Contribution of this paper.}

This paper discusses a novel approach to simultaneous inference that alleviates the need to specify a correlational structure among the marginal errors. In place of the i.i.d. assumption, the proposed approach invokes the assumption of vacuous orientation, a mathematically weaker assumption that delivers logically stronger conclusions.  The {vacuous orientation} assumption retains what the i.i.d. assumption implies on the {configuration} of the measurement errors, i.e. their sum of squares follows a $\chi_k^2$ distribution. On the other hand, it relaxes the implication that the error orientation is isotropic in $\RRR^{K}$.

The proposed model employs the Dempster-Shafer Extended Calculus of Probability (DS-ECP) \cite{dempster2008dempster,dempster2014statistical}, a logical framework for probabilistic reasoning. DS-ECP dictates the specification, combination and processing of marginal information, including observed data, model structure, as well as available (but optional) prior distributional information. The model delivers posterior inference in the form of a probability triple $(\tp, \tq, \tr)$ that sums to one, representing evidence ``for'', ``against'', and ``don't know'' towards a hypothesis one wishes to ascertain. As a generalization to ordinary probability reasoning, DS-ECP is a system of stochastic three-valued logic. The third value augmentation, i.e. a possibly nonzero $\tr$, provides a flexible vocabulary to express partial information, including the complete lack of information. An advantage of DS-ECP over the Bayesian approach is that does not require the modeler to possess {\it a priori} distributional knowledge about unknown parameters, while at the same time enjoying the logical coherence as does the Bayesian approach. 

The DS-ECP analysis framework is a marriage between belief function, an imprecise probability construction, and a functional approach to statistical inference. The model specification in DS-ECP bears resemblance to structural \cite{fraser1968structural} and functional models \cite{dawid1982functional}, as well as the modern approach of generalized fiducial inference \cite{hannig2016generalized}. Other statistical  approaches that leverage non-additive probability measures include robust Bayes \cite{berger1994overview}, inferential models (IM) \cite{martin2015inferential}, and outer probability measures \cite{houssineau2018parameter}. Notably, the concept of validity emphasized by the IM approach in the sense of frequentist coverage is echoed in the analysis of calibration properties in Section~\ref{subsec:rectangle} of this paper. 

The remainder of this paper is organized as follows. Section~\ref{sec:model} lays out the set of weakened assumptions underlying the proposed approach, and demonstrate the combination and projection operations following the inferential recipe of DS-ECP, including a small example to illustrate the recipe itself. Section~\ref{sec:posterior} discusses posterior inference for important types of hypotheses, including linear and quadratic forms, their relationships to hypothesis testing and confidence regions, as well as frequentist coverage properties. In the presence of an exploding number of dependent hypotheses, the posterior probabilities from the vacuous orientation model is uniformly distributed, just as a well-calibrated $p$-value under the null sampling model. In contrast, the Bonferroni approach, which boasts no assumption on the dependence structure of hypotheses, behaves unnecessarily conservatively. Section~\ref{sec:unknown-s} discusses the case in which the error variance $S^2$ is not known but follows a prior distribution. Section~\ref{sec:discussion} discusses potential generalizations to the proposed model.

\section{Model}\label{sec:model}

The proposed model is based on the following state space structure, consisting of\footnote{In general, we require the state space $\Omega$ be the Cartesian product of a collection of marginal spaces, and be endowed with the product topology.}
\begin{equation}
	\left(\hY,\hM,\hE,S^{2}\right) \in \Omega,
\end{equation}
where $\Omega=\Omega_{\hY}\times\Omega_{\hM}\times\Omega_{\hE}\times\Omega_{S^{2}} = \RRR^{3k+1}$.  $\hY$ is a $k$-vector of observable measurements, and $\hM$ the corresponding vector of unknown parameters whose values we wish to learn. $\hE$ is a vector of measurement errors and $S^{2}$ a variance parameter associated with them, both to be defined soon.

\subsection{Marginal evidence.} 

A piece of {marginal evidence} is a mathematical statement regarding subsets of the state space. Observations, together with modeling assumptions, make up the marginal evidence available for analysis. In DS-ECP models, the word {\it margin} may carry a more general meaning than it does in ordinary statistical models, without being confined  to individual dimensions of the state space. A margin can be subsets, as well as (single- or multi-valued) functions of subsets of the state space. This generalization gives the freedom to specify evidence on higher order structures.

The first piece of marginal evidence we specify is the assumption that the true means are observed subject to {additive error}. That is,
\begin{equation}\tag{i}\label{eq:error}
\quad \hY - \hM = \hE.
\end{equation}
The next piece of marginal evidence concerns the {observed data}. That is, the observable measurement $\hY$ realizes to a particular value
\begin{equation}\tag{ii}\label{eq:x}
\quad \hY = \hy.
\end{equation}
While this may look obvious, we spell it out explicitly in anticipation of potentially censored or truncated observations. The third assumption is on the distribution of {\it error configuration}:
\begin{equation}\tag{iii}\label{eq:configuration}
\hE'\hE  = S^2 U,\quad \text{where }U \sim \chi^2_k,
\end{equation}
the Chi-squared distribution with $k$ degrees of freedom. Notice that the sum of squares of the measurement errors is injected with stochastic evidence through the introduction of the variable $U$. It is the modeler's assertion that $U$ be {\it auxiliary}, in the sense that  (\ref{eq:configuration}) is the only way through which the distribution of $U$ injects knowledge into the state space. Besides through (\ref{eq:configuration}), no further information can be learned from $U$ about any aspect of the state space. 

The error configuration assumption says that the sum of squares of the measurement errors follows a scaled $\chi^2$ distribution with a scaling factor $S^2$. As alluded to in Section~\ref{sec:intro}, that the error configuration is distributed as Chi-squared is a necessary but insufficient consequence of the normal i.i.d. assumption.  To see this insufficiency, consider the following probability specification for ${\bf E}$ when the dimension $k \ge 2$: $E_{1}\sim k^{-1/2}\cdot\text{sgn}\left(0.5\right)\cdot\chi_{k}$, where $\text{sgn}\left(0.5\right)$ is a random ``$+/-$'' sign, and $\chi_{k}$ is an independent Chi distribution with $k$ degrees of freedom. Also, $E_{i} = E_{1}$ for all $i=2,\cdots,k$. The implied distribution of ${\bf E}$ asserts that all $E_i$'s are fully correlated. This specification is compliant with the vacuous orientation model, but not with the normal i.i.d. model. Therefore, to replace the i.i.d. assumption with (\ref{eq:configuration}) is a proper weakening of model assumptions. Conclusions derived from the vacuous orientation model are logically stronger.

\begin{figure}
\floatconts{fig:circle}
{\caption{The error configuration assumption (\ref{eq:configuration}) says that the sum of squares of the error terms follows a scaled $\chi^2_k$ distribution. It posits a $k$-sphere centered at $\hM$ with a random radius. Assumption on the error orientation is vacuous, i.e., no knowledge of location on a given $k$-spherical shell is available, contrasting the isotropic orientation assumption implied by the normal i.i.d. model illustrated in Figure~\ref{fig:iid}.}}
{\label{fig:circle}
\includegraphics[width=\linewidth]{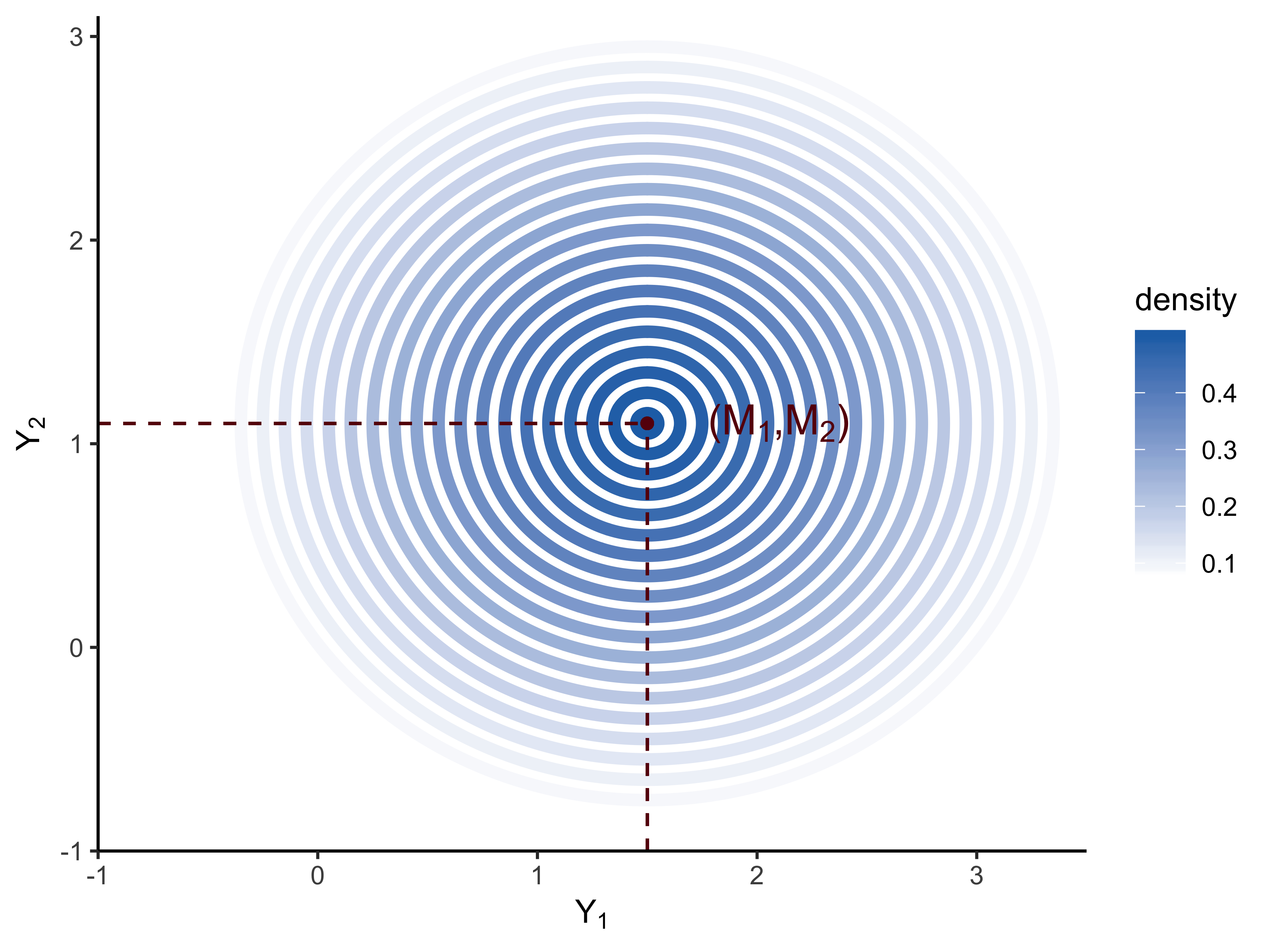}
} 
\end{figure}

The terminology {\it configuration} and {\it orientation} are due to \cite{dempster1969elements}, who used them in a regression context. For  $\hX$ a linear regression design matrix, configuration refers to the matrix product $\hX'\hX$, whereas orientation refers to the information that remains in $\hX$ given $\hX'\hX$. The vacuous orientation assumption refers to the fact that only the marginal distribution of the error configuration is specified, while the error orientation is left unspecified. With this partial specification, the model admits to full ignorance on the correlational structure of the error terms $\hE$. Figure~\ref{fig:circle} is an illustration of the error configuration assumption when $k = 2$. 

We will need a last assumption concerning the variance parameter $S^2$. Consider for now the {\it known variance} case:
\begin{equation}\tag{iv}\label{eq:s-fix}
S^2 = s^2,
\end{equation}
for $s$ a positive real number. In Sections~\ref{sec:model} and~\ref{sec:posterior}, we discuss posterior inference for $\hM$ based on the known variance assumption. The unknown variance case is discussed in Section~\ref{sec:unknown-s}, where we introduce a variant of (\ref{eq:s-fix}) that allows $S^2$ to bear a prior distribution. Reasoning and computation contingent upon either variants are virtually the same. 

\subsection{Independence of marginal evidence.}

As a prerequisite for the next steps of DS-ECP analysis, it is important that the specified collection of marginal evidence  be judged as an {\it independent} body of evidence. Independence justifies the use of Dempster's Rule to combine the marginal information in the joint state space. 

To be concrete, let $\EEE$ denote a collection of $J$ pieces of marginal evidence. $\EEE$ here stands for ``$\EEE$vidence'', to be distinguished from $\hE$ the measurement error. For the model under contemplation, $ \EEE = \{(\ref{eq:error}), (\ref{eq:x}), (\ref{eq:configuration}), (\ref{eq:s-fix})\}$. Let $U_j$ be the auxiliary variable associated with the $j^{th}$ piece of evidence in $\EEE$, which follows a known distribution $\mu_j$. For example, $U$ in (\ref{eq:configuration}) follows the $\chi^2_k$ distribution. Marginal pieces of evidence that are deterministic, such as (\ref{eq:error}), (\ref{eq:x}) and (\ref{eq:s-fix}) which are mere equality statements about margins of the state space, are regarded as associated with constant auxiliary variables. Evidence independence is defined as follows.

\begin{definition}[Independent marginal evidence.]\label{def:independence}
A body of marginal evidence $\EEE$ consisting of $J$ pieces is said to be independent, if the auxiliary variables associated with each piece are all statistically independent. That is, for $U_j \sim \mu_j$ where $j = 1,\cdots, J$, we have that 
\begin{equation}
\left(U_{1},\cdots,U_{J}\right)\sim \mu_{1}\times\cdots\times\mu_{J}.
\end{equation}
\end{definition}

Since constant variables are independent of other variables, all deterministic evidence are naturally independent. By Definition~\ref{def:independence}, the body of evidence for the vacuous orientation model $\EEE = \{(\ref{eq:error}), (\ref{eq:x}), (\ref{eq:configuration}), (\ref{eq:s-fix})\}$ is independent, and is eligible for the next steps of DS-ECP analysis.

Note that the independence of evidence discussed here is distinct from the assumption of independence about the measurement errors. The latter is implied by the normal i.i.d. assumption, precisely what the proposed model attempts to rid by replacing with the mathematically weaker assumption of  $\chi_k^2$ error configuration. Contingent upon the configuration, the errors may or may not be independent. On the other hand, the independence of evidence assumption claims that the information that the error configuration behaves in a certain way is independent of other pieces of information, such as that the errors are additive, the variance is known, and so on. In other words, independence of evidence refers to statistical independence of the auxiliary variables, a separate notion from (in)dependence among margins of the state space. Independence of evidence is just as much a subjective judgment on the part of the modeler. In complex models, through careful specification using a join tree structure \cite{kong1987multivariate}, one can successively construct independent bodies of evidence such that they're eligible for DS-ECP analysis using Dempster's Rule of Combination.

\subsection{Evidence projection and combination.}

The processing of marginal evidence consists of three steps: 1) {up-projection} of marginal evidence to the joint state space, 2) {combination} of evidence in the joint state space, and 3) {down-projection} of combined evidence to margins of interest. The three steps are explicated in this section.

\subsubsection{Up-projection of marginal evidence.}

To up-project a piece of marginally specified evidence is to extend its evidence statement involving only a subset of the state space variables, into one that concerns the entire state space. Algebraic relationships defined on margins of the state space  are extended into cylinder sets, spanning the remainder dimensions that were not mentioned in the evidence statement. For example, (\ref{eq:error}) implies the following partitioning of $\Omega$:
\begin{equation}
	\{\left(\hY,\hM,\hE,S^{2}\right) \in \Omega : \hY - \hM = \hE \},
\end{equation}
and similarly for statement (\ref{eq:configuration}),
\begin{equation}
	\{\left(\hY,\hM,\hE,S^{2}\right) \in \Omega : \hE'\hE = S^2 U \}.
\end{equation}
In other words, the up-projection process levels the playground for all marginal evidence, such that they become comparable statements concerning the same joint space $\Omega$, and ready to be combined there.

\subsubsection{Evidence combination via Dempster's Rule.}

Since $\EEE = \{ (\ref{eq:error}), (\ref{eq:x}), (\ref{eq:configuration}), (\ref{eq:s-fix})\}$ is judged to be independent, we apply Dempster's rule to combine its component pieces, that is, by taking the intersections of all up-projected evidence whenever they are nonempty. These nonempty intersections become the new focal sets, representing the combined evidence from $\EEE$. In this model, the combination of $\EEE$ implies a class of subsets of $\Omega$ of the following structure
\begin{multline}\label{eq:RI1}
	\RR_{\EEE} \eqdef \{\left(\hY,\hM,\hE,S^{2}\right) \in \Omega : \\ \hY = \hy,\, \hY - \hM = \hE,\, \hE'\hE = S^2 U,\, S^2 = s^2 \},
\end{multline}
where $U \sim \chi^2_k$. Notice that $\RR_{\EEE}$ is a multi-valued map \cite{dempster1967upper} from $U$ to subsets of $\Omega$. Since $U$ bears a known distribution, $\RR_\EEE$ can be regarded as a random subset of $\Omega$ whose distribution is inherited from that of $U$. The probability density function of $U$ dictates the mass function of $\RR_{\EEE}$. 

Upon combining arbitrary body of evidence, the auxiliary variable distribution may need to be revised. To be precise, the domain on which the auxiliary variable is {\it a priori} defined may be reduced, to exclude those values that result in algebraic incompatibility among pieces of marginal evidence. These incompatible values are those that correspond to marginal focal sets that result in empty intersections with other marginal focal sets. Empty intersections are eliminated from the combination process, while the weights of the remainder non-empty intersections renormalize to one. The revision of the auxiliary distribution proceeds as follows.

Denote $\mu$ the prior probability of the auxiliary variable $\hU$, associated with a body of evidence $\EEE$ and measurable with respect to $\sigma\left({\Xi}\right)$. Upon combining $\EEE$, $\mu$ is revised to $\mu_{\EEE}$, the posterior probability measurable with respect to $\sigma\left({\Xi}_\EEE\right) \subset \sigma\left({\Xi}\right)$, where $\Xi_{\EEE}=\left\{ u\in\Xi:\RR_{\EEE}\left(u\right)\neq\emptyset\right\}$, and
\begin{equation}\label{eq:auxiliary-measure}
\mu_{\EEE}=\left(\mu\times{\bf 1}_{\Xi_{\EEE}}\right)/\mu\left(\Xi_{\EEE}\right), 
\end{equation}
where the indicator function ${\bf 1}_{A}(S) = 1$ if $S \subseteq A$ and $0$ otherwise. In case the denominator $\mu\left(\Xi_{\EEE}\right)$ is $0$, (\ref{eq:auxiliary-measure}) may be alternatively defined via regular conditional probability or limiting arguments. For the case at hand, it just so happens that none of the four component evidence of $\EEE = \{ (\ref{eq:error}), (\ref{eq:x}), (\ref{eq:configuration}), (\ref{eq:s-fix})\}$ raises algebraic conflict for any given value of the auxiliary variable $U$. Thus, the revision of auxiliary distribution is trivial, namely $\mu_{\EEE} = \mu$ which is still the $\chi_k^2$ distribution.

\subsubsection{Down-projection to margins of interest.}
Rarely is the case that we wish to draw inference about the entire state space. Often, we are only interested in a particular margin of the state space, such as $\hM$ the parameter of interest. To reduce computational burden, the random subset $\RR_{\EEE}$ defined on $\Omega$ is projected onto the margin of interest $\Omega_\hM$. This process is called down-projection. The projection of $\RR_{\EEE}$ onto $\Omega_\hM$ is 
\begin{equation}
\RRM{} \eqdef \{\hM\in\Omega_{\hM}:\left(\hM-\hy\right)'\left(\hM-\hy\right)=s^{2}U\}
\end{equation}
where $U \sim \mu_\EEE$,  the $\chi_k^2$ distribution. $\RRM{}$ is again a random subset of $\Omega_\hM$ whose distribution is dictated by $U$. For every realization $U = u$, $\RRM{}\left(u\right)$ is a $k$-sphere centered at $\hy$ with radius $s\sqrt{u}$. Repeated draws of $U$ following the $\chi^2_k$ distribution result in a collection of concentric $k$-spheres of varying radii. The random $k$-sphere $\RRM{}$ embodies posterior inference for $\hM$. Section~\ref{sec:posterior} discusses posterior inference based on $\RRM{}$,  expanding on its properties in greater detail.

\subsubsection{Projection and combination: an example.}

We given an example to illustrate the projection and combination operations discussed above. Let $\hX=\left(X_{1},X_{2}\right)$ be two independent tosses of a same coin, whose chance of landing as head is $\theta$. Suppose the following Bernoulli experiment for $i=1,2$: 
\begin{equation}
X_{i}={\bf 1}\left(U_{i}\le\theta\right),	
\end{equation}
where the auxiliary variables $U_{i}\overset{\text{iid}}{\sim}\mu_i = \text{Unif}\left(0,1\right)$. A head followed by a tail was observed: $X_1 = 1$ and $X_2 = 0$. Gathering the above evidence as $\mathbb{E}'$, we have that
\begin{equation}
\RR_{\EEE'}=\{\left(\hX,\theta,\hU\right):\hX=(1,0),X_{i}={\bf 1}\left(U_{i}\le\theta\right)\},
\end{equation}
where the prior auxiliary variable distribution $\mu=\mu_{1}\times\mu_{2}$ is uniform on the unit square. Notice that the data implies $U_{1}\le\theta$ and $U_{2}>\theta$, hence for any given $\theta$, $U_{2}$ must be greater than $U_{1}$. This restricts the domain of the auxiliary variable to $\Xi_{\EEE'}=\left\{ {\bf u}\in\left[0,1\right]^{2}:u_{1} < u_{2}\right\}$, the northwest triangle of the unit square. It follows from (\ref{eq:auxiliary-measure}) that the revised auxiliary variable distribution $\mu_{\EEE'} \propto {\bf 1}_{\Xi_{\EEE'}}$, uniform over the northwest triangle of the unit square, which is equivalent to the probability distribution induced by a pair of Uniform order statistics. Down-projecting $\RR_{\EEE'}$ to the $\theta$ margin, we obtain the random subset that embodies posterior inference for $\theta$:
\begin{equation}
\RR_{\theta \mid \EEE'}=\{\theta \in [0,1] :U_{1}<\theta\le U_{2}\},
\end{equation}
where ${\bf U}\sim \mu_{\EEE'}$. $\RR_{\theta \mid \EEE'}$ is a half-closed, half-open random interval on $[0,1]$. Its left and right end points are marginally distributed as $Beta(1,2)$ and $Beta(2,1)$ respectively.

\section{Posterior inference}\label{sec:posterior}

Posterior inference about unknown quantities in the state space are expressed through a probability triple $\left(\tp, \tq, \tr\right)$, representing weights of evidence ``for'', ``against'', and ``don't know'' about an assertion concerning subsets of the state space. When $\RRM{}$ is the down-projected random subset that embodies posterior inference for $\hM$, it inherits randomness from the revised auxiliary variable $\hU \sim \mu_\EEE$. Define a trio of set functions $\tp$, $\tq$, $\tr: \Omega_{\hM} \to [0, 1]$ such that for all $H \in \sigma\left(\Omega_{\hM}\right)$,
\begin{align}
\tp\left(H \right)	&=	\int_{\left\{ u\in\Xi_{\EEE}:\RRM{}\left(u\right)\subseteq H\right\} }d\mu_{\EEE}, \label{eq:tp} \\
\tq\left(H\right)	&=	\int_{\left\{ u\in\Xi_{\EEE}:\RRM{}\left(u\right)\subseteq H^c\right\} }d\mu_{\EEE} = \tp\left(H^{c}\right), \label{eq:tq}\\
\tr\left(H\right)	&=	1-\tp\left(H\right)-\tq\left(H\right), \label{eq:tr}
\end{align}
with $\tp + \tq + \tr = 1$. Note that $\left(\tp, \tq, \tr\right)$ are implicit functions of $\EEE$, a dependence we suppress for notational simplicity. The $\left(\tp, \tq, \tr\right)$ representation of posterior inference is an alternative to using a pair of belief and plausibility functions \cite{shafer1976mathematical}. In particular, $\tp$ is a belief function on $\Omega_{\hM}$. $1 - \tq$, or equivalently $\tp + \tr$, is its conjugate plausibility function. The three-valued representation has the advantage that it explicitly acknowledges a possibly non-zero $\tr$, the ``don't know'' probability which reflects the extent of structural uncertainty within the model.

\subsection{Linear forms and hypothesis tests.}

Let $\hC$ be a real-valued $p$ by $k$ matrix, where $p$ can be smaller than, the same as, or larger than $k$. A consistent system of equations 
\begin{equation}\label{eq:cma}
	\hC\hM = \ha
\end{equation}
is a linear margin of the parameter space $\Omega_\hM$ of dimension $k - \text{rank}(\hC)$. Inference about linear margins of the parameter space is the most common type of posterior inference, encompassing a variety types of hypotheses.

We are interested in drawing inference about a linear margin in the form of (\ref{eq:cma}). Define the summary statistic 
\begin{equation}\label{eq:tx}
\tx =  \left(\ha - \hC\hy \right)'\left(\hC\hC'\right)^{-1}\left(\ha - \hC\hy \right),
\end{equation}
where in case $p > \text{rank}(\hC)$, the inverse is defined as the Moore-Penrose pseudoinverse. Results below concern posterior inference for one- and two-sided linear forms of $\hM$, expressed in terms of $\tx$.

\begin{theorem}[Two-sided linear form]\label{thm:2side}
For a two-sided linear hypothesis $H: \hC\hM = \ha$, The DS posterior probabilities concerning $H$ are
\begin{equation}
\{ \tp\left(H\right), \tq\left(H\right),\tr\left(H\right) \} = \{ 0, F\left(\tx\right), 1-F\left(\tx\right) \}
\end{equation}
where $F$ is the CDF of the scaled $\chi^2_k$ distribution with scaling factor $s^2$.
\end{theorem}

Proof of Theorem~\ref{thm:2side} recognizes that $\tx$ is the minimum square radius of $k$-spheres of the form $\RRM{}$ to intersect with the linear subspace $\hC\hM = \ha$. The probability that $\RRM{}$ does not intersect with the linear subspace contributes to $\tq(H)$, and the probability that it does contributes to $\tr(H)$.

\begin{theorem}[One-sided linear form]\label{thm:1side}
For a one-sided linear hypothesis $H: \hC\hM \le \ha$, The DS posterior probabilities concerning $H$ are
\begin{equation}
\{ \tp\left(H\right), \tq\left(H\right),\tr\left(H\right) \} =
\{ F\left(\tx\right), 0, 1-F\left(\tx\right) \}
\end{equation}
if $\hC\hy \le \ha$, and 
\begin{equation}\label{eq:2side-up}
\{ \tp\left(H\right), \tq\left(H\right),\tr\left(H\right) \} =
\{ 0, F\left(\tx\right), 1-F\left(\tx\right) \}
\end{equation}
otherwise. $F$ is defined as in Theorem~\ref{thm:2side}.
\end{theorem}

The hypothesis in Theorem~\ref{thm:1side} is a halfspace formed by the hyperplane which Theorem~\ref{thm:2side} posits as hypothesis. Depending on $\hy$, the location of the center of $\RRM{}$, the hypothesized halfspace may or may not contain $\RRM{}$ with positive probability. If $H$ is not supported at face value by empirical evidence, that is, either $H$ is degenerate relative to $\Omega_\hM$ or it asserts to the contrary of what the observation appears to be (e.g. $H: \hC\hM \le \ha$ but $\hC\hy \lneq
  \ha$), then $\tp(H)$, the probability ``for'' $H$, is zero. On the other hand, if $H$ is supported by empirical evidence, then $\tq(H)$, or the probability ``against'' $H$, is zero. The probability of ``don't know'', $\tr(H)$, is the same as long as $H$ concerns the same linear subspace in $\Omega_\hM$. This again demonstrates the intuitive appeal of $\tr$ as a posterior summary statistic: its value is reflective of the inherent structural uncertainty regarding the hypothesis with respect to the collection of evidence. That is, it reflects the extent to which $\EEE$ is able to discern anything about $H$ at all, while  agnostic to the direction of support for it. 
 
Examples below showcase a variety of hypothesis types to which the above theorems are applicable. Posterior probabilities are examined in comparison to the frequentist answers under the normal i.i.d. model.

\begin{example}[test for all means]\label{ex:all0}
Suppose the null hypothesis is $H:\hM = {\bf 0}$, and the alternative $H^{c}:\hM \neq {\bf 0}$. $\hC$ is the identity matrix, and $\ha = {\bf 0}$. The test statistic simplifies to $\tx = \sum_{i=1}^{k} y_{i}^{2}$. The posterior probability ``for'' $H$ is $0$, and posterior probability of ``don't know'' about $H$ is 
\begin{equation}\label{eq:up-all0}
\tr\left(H\right) = \Gamma\left(k/2,\sum y_{i}^{2}/2s^{2}\right)/\Gamma\left(k/2\right),
\end{equation}
which is the survival probability (i.e. one minus the cumulative probability) of a $\chi_k^2$ distribution evaluated at $\sum y_{i}^{2}/s^{2}$.
\end{example}

Notice that (\ref{eq:up-all0}) is identical to the $p$-value one would obtain under the i.i.d. sampling model, for which the likelihood ratio test is based on the same test statistic. It has an exact reference distribution of $\chi_k^2$ under the null. Example~\ref{ex:one0} to follow seems like a slight modification from Example~\ref{ex:all0}, but the solution it supplies can be rather distinct.

\begin{example}[test for one mean]\label{ex:one0}
Suppose the null hypothesis is $H:M_1 = 0$, and the alternative $H^{c}:M_1 \neq 0$. Here, $\hC$ is a column vector of $0$'s at all except the first entry, which takes the value of $1$. The test statistic is $\tx = y_{1}^{2}$.   The posterior probability ``for'' $H$ is $0$, and posterior probability of ``don't know'' about $H$ is
\begin{equation}\label{eq:up-one0}
\tr\left(H\right)	
	 = \Gamma\left(n/2,y_{1}^{2}/2s^{2}\right)/\Gamma\left(n/2\right),
\end{equation}
which is the survival probability of a $\chi_k^2$ distribution evaluated at $y_{1}^{2}/s^{2}$.	
\end{example}

When $k = 1$, Example~\ref{ex:one0} reduces to Example~\ref{ex:all0}, and (\ref{eq:up-one0}) agrees with the $p$-value obtained under the i.i.d. sampling model using the likelihood ratio test. However when $k > 1$, (\ref{eq:up-one0}) is larger than the $p$-value obtained under the i.i.d. sampling model, which utilizes $\chi_1^2$ as the reference distribution for $\tx$ regardless of $k$. Under the vacuous orientation assumption, the reference distribution $\chi_k^2$ grows with $k$, and is more conservative than the i.i.d. model.

\begin{example}[test for all pairwise contrasts]\label{ex:contrast}
Suppose we conduct a simultaneous test for the null hypothesis that all pairwise means are identical. That is, 
\begin{equation*}
H=\cap_{1\le i<j\le k}H_{i,j},\quad H_{i,j}:M_{i}=M_{j}
\end{equation*}
The alternative hypothesis $H^{c}$ is that at least one of the equalities doesn't hold. $\hC$ is a $k(k-1)/2$ by $k$ matrix of all pairwise contrasts with $\text{rank}(\hC) = k - 1$, and $\ha = {\bf 0}$. The test statistic simplifies to 
\begin{equation}
\tx=\sum_{i}\left(\overline{y}-y_{i}\right)^{2}-\frac{1}{k}\sum_{i,j}\left(\overline{y}-y_{i}\right)\left(\overline{y}-y_{j}\right).	
\end{equation}
The posterior probability ``for'' $H$ is zero, and that of ``don't know'' about $H$ can be found via (\ref{eq:2side-up}) with the reference distribution again a $s^2$-scaled $\chi_k^2$ distribution. 

\begin{figure}
\floatconts{fig:pvalue-contrast}
{\caption{Distribution of $\tr(H)$ for all pairwise contrasts under the null sampling model for various $k$. As the dimension $k$ increases, the distribution of $\tr(H)$ approaches uniform, which is the distribution of a correctly calibrated $p$-value.}}
{
\label{fig:pvalue-contrast}
\includegraphics[width=\linewidth]{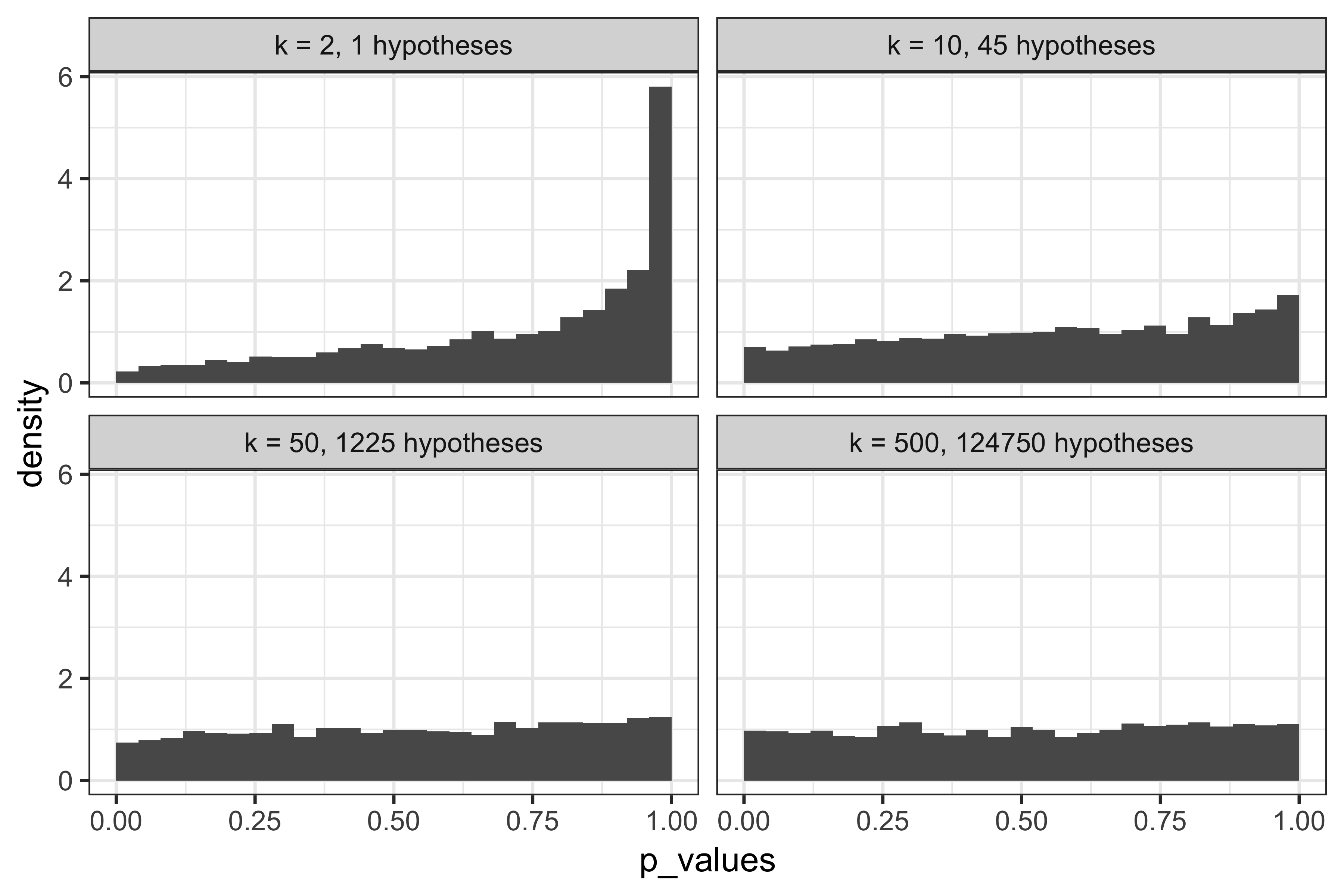}
} 
\end{figure}
\end{example}

Figure~\ref{fig:pvalue-contrast} displays values of $\tr(H)$ for simulations under the null i.i.d. sampling model for various $k$. The number of pairwise contrasts tested is on quadratic order of $k$, but the compound hypothesis $H$ always spans a $1$-dimensional subspace of $\Omega_\hM$.  As $k$ increases, the distribution of $\tr(H)$ using simulated null data approaches the uniform distribution, which is the distribution of a correctly calibrated $p$-value under the null model. This stands in sharp contrast to the Bonferroni procedure. Observations obtained from a similar simulation experiment shows that, while the Bonferroni correction controls for the the overall test size $\alpha$,  for larger $k$ it becomes increasingly conservative, in that the fraction of null samples leading to a rejection of $H$ is far less than $\alpha$ for all $\alpha$ in range. The culprit to Bonferroni's conservatism is that it adjusts the individual test size by mindlessly dividing  $\alpha$ with the number of hypotheses tested. It neither respects nor utilizes the logical connection among the large number of hypotheses. The proposed model captures such feature, and delivers logically coherent posterior inference reflective of the geometry of the hypothesis space. 

\subsection{Quadratic forms and calibrated credible regions.}

The sampling distribution under the i.i.d. assumption is the spherical multivariate normal distribution. The associated confidence region for its mean vector is thus spherical. The proposed model $\EEE$ induces random subsets in the parameter space in the form of concentric spherical shells. In this section, we consider credible regions for $\hM$ that are of quadratic forms. These credible regions are special in that they deliver {\it sharp} posterior inference that are also {\it calibrated} with respect to the normal i.i.d. model. 

\begin{definition}[sharp inference]\label{def:sharp}
Given a body of marginal evidence $\EEE$,  we say that the posterior inference for $A \in \sigma(\Omega_\hM)$ is sharp if $\tr(A) = 0$.	
\end{definition}

\begin{definition}[calibrated inference]
Given $\EEE$, we say that posterior inference is lower-calibrated for $A \in \sigma(\Omega_\hM)$ with respect to a sampling model $\hY \mid \hM^* \sim P^*$, if the posterior probability ``for'' $A$ is equal to the frequentist coverage probability of $A$ under the sampling model, when $A$ is viewed as a function of $\hY$. That is,
\begin{equation}
	\tp(A)=P^{*}\left(\hM^* \in A \right).
\end{equation}
Similarly, posterior inference for $A$ is upper-calibrated if 
\begin{equation}
	\tq(A)=P^{*}\left(\hM^* \in A^c \right).
\end{equation}
Posterior inference for $A$ is calibrated if it is both lower- (or upper-) calibrated and sharp.
\end{definition}

For $\alpha \in [0, 1]$, define the $(1-\alpha)$ posterior credible region
\begin{equation}\label{eq:credible_A}
A_{\alpha}=\left\{ \hM\in\Omega_{\hM}:\left(\hM-\hy\right)'\left(\hM-\hy\right)\le F^{-1}_{1-\alpha}\right\},
\end{equation}
where $F^{-1}_\alpha$ is the $\alpha^{th}$-quantile of the posterior auxiliary distribution $\mu_\EEE$.  By (\ref{eq:tp}) and (\ref{eq:tq}), we have that
\begin{equation}\label{eq:1-alpha}
	\tp(A_\alpha) = 1-\alpha, \quad \tq(A_\alpha) = \alpha.
\end{equation}
The posterior inference coincides with the i.i.d. model inference, when hypotheses of quadratic forms (such as $A_\alpha$) are contemplated. This should come as no surprise, since when $A_\alpha$ is viewed as a quadratic function of $\hy$, its probabilistic property is precisely the aspect of the i.i.d. assumption that is preserved under the vacuous orientation assumption. 

\begin{theorem}[sharp credible region.]\label{thm:sharp}
$A_{\alpha}$ is a sharp posterior credible region for $\hM$. That is, $\tr(A_{\alpha}) = 0$.
\end{theorem}

Theorems~\ref{thm:sharp} is a direct consequence of (\ref{eq:1-alpha}). We also have the following.

\begin{theorem}[calibrated credible region.]\label{thm:calibrated}
$A_{\alpha}$ is a calibrated posterior credible region with respect to the i.i.d. sampling model, for all $\hM^*$ and all $\alpha$. 
\end{theorem}

To prove Theorem~\ref{thm:calibrated}, one just need to show that $\alpha=P^{*}\left(\left(\hY-\hM^{*}\right)'\left(\hY-\hM^{*}\right)>F_{1-\alpha}^{-1}\right)$ for all $\hM^*$ and all $\alpha$. That indeed is the case, when $P^{*}$ is the normal i.i.d. sampling distribution specified in (\ref{eq:sampling}) and (\ref{eq:indep}).

\subsection{Rectangular parallelepipedal regions.}\label{subsec:rectangle}

Under the sampling model in (\ref{eq:sampling}), for $i = 1,\cdots, k$, a size-$\alpha$ test for hypothesis $H_{i}:M_{i}=0$ is dual to a confidence interval for $M_i$ of the form $\left(y_{i}\pm\Phi\left(1-{\alpha}/{2}\right)\cdot s\right)$. If a compound null hypothesis $H:\cap_{i=1}^{k}H_{i}$ is contemplated, and if the confidence intervals for each component hypothesis are calculated as above, the familywise error rate (i.e. the probability of making at least one false rejection) exceeds $\alpha$. As alluded to in Section~\ref{sec:intro}, the Bonferroni procedure accounts for the fact that $k$ hypotheses are tested simultaneously. To maintain the familywise error rate at no more than $\alpha$, the test size for each component hypothesis is reduced to $\alpha/k$. Hence, a Bonferroni-corrected test for hypothesis $H$ is dual to the rectangular confidence region
\begin{equation}\label{eq:bonf-rectangle}
\otimes_{i=1}^{k}\left(y_{i}\pm b_{\alpha}\cdot s\right),	
\end{equation}
where $b_{\alpha}=\Phi\left(1-{\alpha}/{2k}\right)$ is the Bonferroni-corrected and standardized half width of the univariate interval.

To parallel the Bonferroni confidence region, consider rectangular parallelepipedal regions of the form
\begin{equation}\label{eq:rectangle}
C_{\alpha} = \left\{ \hM\in\Omega_{\hM}:\hM\in\otimes_{i=1}^{k}\left(y_{i}\pm c_{\alpha}\cdot s\right)\right\}.
\end{equation}
Posterior probabilities associated with $C_\alpha$ can be regarded as a function of the standardized half width $c_{\alpha}$. We have the following results.
 
 \begin{theorem}\label{thm:rectangle}
 The posterior probabilities of $C_{\alpha}$ are
\begin{multline}
\{ \tp\left(C_{\alpha} \right) , \tq\left(C_{\alpha} \right) , \tr\left(C_{\alpha} \right) \} = \\	\{ F\left(c_{\alpha}^{2}\right), 	F\left(kc_{\alpha}^{2}\right) - F\left(c_{\alpha}^{2}\right), 1 - F\left(kc_{\alpha}^{2}\right) \}
\end{multline}
where $F$ is the CDF of the $\chi^2_k$ distribution.
\end{theorem}

To prove Theorem~\ref{thm:rectangle}, notice that $\tp\left(C_{\alpha} \right)$ is the probability that $\RRM{}$ is fully contained in $C_\alpha$, i.e. its radius less than or equal to $c_\alpha \cdot s$. On the other hand, $\tq\left(C_{\alpha} \right)$ is the probability that $\RRM{}$ fully contains $C_\alpha$, i.e. its radius greater than $\sqrt{k} c_\alpha\cdot s$. A lemma immediately follows.

\begin{lemma}
Let $\underline{c}_{\alpha}$ and $\overline{c}_{\alpha}$ be the standardized half widths of $C_\alpha$ such that it is lower-calibrated (i.e. $\tp\left(C_{\alpha}\right) = {1-\alpha}$) and upper-calibrated (i.e. $\tq\left(C_{\alpha}\right) = {\alpha}$), respectively. We have
\begin{equation}
\underline{c}_{\alpha} = \left(F^{-1}_{1-\alpha}\right)^{1/2}, \quad	
\overline{c}_{\alpha} =	\left( F^{-1}_{1-{\alpha}}/{k}\right)^{1/2}
\end{equation}
where $F^{-1}_\alpha$ is the $\alpha^{th}$-quantile of the $\chi^2_k$ distribution.
\end{lemma}

Just like the Bonferroni half width $b_{\alpha}$, $\underline{c}_{\alpha}$ is an increasing function of the dimension $k$. On the other hand, $\overline{c}_{\alpha}$ is a decreasing function of $k$. The three quantities are identical when $k = 1$. Table~\ref{tab:c005} displays a comparison among the three quantities for various $k$ at $\alpha = 0.05$.

\begin{table}[htbp]
\floatconts{tab:c005}%
  {\caption{Half widths of $C_{\alpha}$ for various $k$, $\alpha = 0.05$. $\overline{c}_{\alpha} = b_{\alpha} = \underline{c}_{\alpha}$ when $k = 1$. $\overline{c}_{\alpha}<b_{\alpha}<\underline{c}_{\alpha}$ when $k > 1$.}}%
  {\label{tab:c005}
  \begin{tabular}{lrrr}
  \toprule
  $k$ & $\overline{c}_{0.05}$ & ${b}_{0.05}$ & $\underline{c}_{0.05}$  \\ 
  \midrule
  1 & 1.96 & 1.96 & 1.96 \\ 
    2 & 1.73 & 2.24 & 2.45 \\ 
    5 & 1.49 & 2.58 & 3.33 \\ 
   10 & 1.35 & 2.81 & 4.28 \\ 
  100 & 1.12 & 3.48 & 11.15 \\ 
   \bottomrule
\end{tabular}
   } 
\end{table}

We also consider the posterior probabilities for rectangles to which the Bonferroni procedure assigns $1-\alpha$ confidence. A comparison for various $k$ and $\alpha$ is displayed in Table~\ref{tab:bonf}.  As $k$ increases, both the posterior ``for'' and ``against'' probabilities approach $0$, and the posterior ``don't know'' probability approaches $1$. Unlike spherical regions, posterior inference for rectangular regions is not sharp  under the vacuous orientation assumption. This reveals that, to make sharp probabilistic statements on rectangular regions require substantial model input concerning the dependence structure of observation errors. The growing extent of ``don't know'' posterior probability quantifies the extent to which a sharp probabilistic assignment on a rectangular region need to depend on such assumptions, increasingly so as the dimension of the parameter space grows.

\begin{table}[htbp]\label{tab:bonf}
\floatconts{tab:bonf}%
  {\caption{Posterior probabilities associated with Bonferroni $(1-\alpha)$ rectangles (\ref{eq:bonf-rectangle}), for $\alpha = 0.05$ (left) and $0.2$ (right).}}%
  {\label{tab:bonf} 
  \begin{tabular}{lrrr|rrr}
  \toprule
    & \multicolumn{3}{c}{$\alpha = 0.05$} &
      \multicolumn{3}{c}{$\alpha = 0.2$ }  \\ 
  \midrule%{2-5}
$k$ & $\tp$ & $\tq$ & $\tr$ & $\tp$ & $\tq$ & $\tr$  \\ 
  \midrule  
    1 & 0.95 & 0.05 & 0.00 & 0.80 & 0.20 & 0.00 \\ 
    2 & 0.92 & 0.01 & 0.07 & 0.74 & 0.07 & 0.19 \\ 
    5 & 0.75 & 0.00 & 0.25 & 0.48 & 0.00 & 0.52 \\ 
   10 & 0.36 & 0.00 & 0.64 & 0.14 & 0.00 & 0.86 \\ 
  100 & 0.00 & 0.00 & 1.00 & 0.00 & 0.00 & 1.00 \\ 
   \bottomrule
\end{tabular}
} 
\end{table}

\section{Case with unknown variance}\label{sec:unknown-s}

As alluded to in Section~\ref{sec:intro}, under the sampling model in (\ref{eq:sampling}) every unknown parameter $M_i$ is measured only once. There is no hope to extract knowledge about the variance parameter $S^2$ from the data. If $S^2$ is not known precisely, it needs to be assumed to follow a prior distribution. This section describes a modification to the proposed model, to accommodate the case that $S^2$ is not known but rather follow a prior distribution. The {\it unknown variance} assumption posits that
\begin{equation}\tag{iv.2}\label{eq:s-prior}
S^2 = U_{s}, %\; \text{where } U_{s} \sim Q_s 
\end{equation}
where $U_s$ is an auxiliary variable bearing some known prior distribution for $S^2$, and is independent of the auxiliary variable $U$ from (\ref{eq:configuration}). Write $\hU = \left(U, U_s\right)$, and denote by $\EEE.2 = \{ (\ref{eq:error}), (\ref{eq:x}), (\ref{eq:configuration}), (\ref{eq:s-prior})\}$ the body of evidence using the alternative unknown variance assumption.  By assuming independence between $U$ and $U_s$, $\EEE.2$ is judged independent according to Definition~\ref{def:independence}, and is eligible for combination via Dempster's Rule.  The projection and combination operations follow the same fashion as described in Section~\ref{sec:posterior}. $\EEE.2$ implies the following class of subsets of $\Omega$:
\begin{multline}
	\RR_{\EEE.2} = \{\left(\hY,\hM,\hE,S^{2}\right) \in \Omega : \\ \hY = \hy,\, \hY - \hM = \hE,\, \hE'\hE =  S^2 U,\, S^2 = U_s \},
\end{multline}
and the down projection of $\RR_{\EEE.2}$ onto $\Omega_\hM$ is 
\begin{equation}
\RRM{.2} = \{\hM\in\Omega_{\hM}:\left(\hM-\hy\right)'\left(\hM-\hy\right)=U/U_{s} \}.\end{equation}
One can use any preferred distribution as the prior distribution for $S^2$. One computationally convenient choice is the $\text{Inv-}\chi^2$ distribution with $\nu$ degrees of freedom. By independence of $U$ and $U_s$, the ratio $U/U_{s} \sim \mu_{\EEE.2} = \left(k/\nu\right)\FF_{\left(k,\nu\right)}$, the scaled $\FF$-distribution with scaling factor $k/\nu$ and degrees of freedom $k$ and $\nu$. All calculations in Section~\ref{sec:posterior} remain the same after  replacing $\mu_\EEE$ with $\mu_{\EEE.2}$. Posterior inference based on other prior specifications follows the same logic.

\section{Discussion}\label{sec:discussion}

Several useful generalizations to the proposed model are to be explored in future studies.

\subparagraph{Elliptical distributions.} The sampling distribution employed by the i.i.d. model is a multivariate normal distribution with identity covariance matrix, which is a special kind of elliptical distribution. The vacuous orientation assumption can be generalized to other families of elliptical distributions with covariance $\Sigma$, for which the probability distribution is a function of the configuration $\left(\hM-\hY\right)'\Sigma^{-1}\left(\hM-\hY\right)$. Of particular interest are the multivariate $t$ and Laplace distributions, suitable for measurement errors are expected to be both heavy-tailed and dependent.

\subparagraph{Multivariate regression.}

The current model setting can be generalized to accommodate covariate information. Specifically, let $\hM = \hX \hbeta$, where $\hX$ is a $k$ by $p$ design matrix and $\hbeta$ a $p$-vector of coefficients. A vacuous orientation assumption on the observation errors now induces posterior inference about $\hbeta$ in a similarly vacuous manner. Note that the current model produces meaningful simultaneous posterior inference when the number of hypotheses far exceeds the number of unknown parameters. In the same way, the regression model with the vacuous orientation assumption can deliver ``large $p$, small $k$'' inference, that is, when the regression model itself is underdetermined. 

\subparagraph{Finer variance decomposition.}
Lastly, the vacuous orientation assumption itself is also subject to extension. Notice that the configuration of $k$ i.i.d. normal error terms can be decomposed into a collection of up to $k$ variance components, such that they're all independent among each other and with degrees of freedom that sum to $k$. In fact, both the i.i.d. model and the vacuous orientation model are special cases of such variance decomposition. The former is the trivial decomposition into $k$ components, each of degrees of freedom 1. The latter is the canonical decomposition into two components, configuration and orientation, of degrees of freedom 1 and $k-1$ respectively. One can imagine generalizing the vacuous assumption onto other variance decompositions to express degrees of ignorance over the dependence structure among errors.

%% For papers, sections below this line do not count towards the page limit

\acks{The author thanks Professor Arthur P. Dempster for inspiring this work and for valuable discussions, as well as three anonymous reviewers for helpful suggestions.}

\bibliography{gong19.bib}

\begin{thebibliography}{15}
\providecommand{\natexlab}[1]{#1}
\providecommand{\url}[1]{\texttt{#1}}
\expandafter\ifx\csname urlstyle\endcsname\relax
  \providecommand{\doi}[1]{doi: #1}\else
  \providecommand{\doi}{doi: \begingroup \urlstyle{rm}\Url}\fi

\bibitem[Benjamini and Hochberg(1995)]{benjamini1995controlling}
Yoav Benjamini and Yosef Hochberg.
\newblock Controlling the false discovery rate: a practical and powerful
  approach to multiple testing.
\newblock \emph{Journal of the Royal statistical society: series B
  (Methodological)}, 57\penalty0 (1):\penalty0 289--300, 1995.

\bibitem[Benjamini and Yekutieli(2001)]{benjamini2001control}
Yoav Benjamini and Daniel Yekutieli.
\newblock The control of the false discovery rate in multiple testing under
  dependency.
\newblock \emph{The annals of statistics}, 29\penalty0 (4):\penalty0
  1165--1188, 2001.

\bibitem[Berger(1994)]{berger1994overview}
James~O Berger.
\newblock An overview of robust {Bayesian} analysis.
\newblock \emph{Test}, 3\penalty0 (1):\penalty0 5--124, 1994.

\bibitem[Dawid and Stone(1982)]{dawid1982functional}
Philip~A Dawid and Mervyn Stone.
\newblock The functional-model basis of fiducial inference.
\newblock \emph{The Annals of Statistics}, 10\penalty0 (4):\penalty0
  1054--1067, 1982.

\bibitem[Dempster(1967)]{dempster1967upper}
Arthur~P Dempster.
\newblock Upper and lower probabilities induced by a multivalued mapping.
\newblock \emph{The Annals of Mathematical Statistics}, 38\penalty0
  (2):\penalty0 325--339, 1967.

\bibitem[Dempster(1969)]{dempster1969elements}
Arthur~P Dempster.
\newblock \emph{Elements of continuous multivariate analysis}.
\newblock Addison-Wesley, 1969.

\bibitem[Dempster(2008)]{dempster2008dempster}
Arthur~P Dempster.
\newblock The {Dempster-Shafer} calculus for statisticians.
\newblock \emph{International Journal of Approximate Reasoning}, 48\penalty0
  (2):\penalty0 365--377, 2008.

\bibitem[Dempster(2014)]{dempster2014statistical}
Arthur~P Dempster.
\newblock Statistical inference from a dempster--shafer perspective.
\newblock \emph{Past, Present, and Future of Statistical Science}, pages
  275--288, 2014.

\bibitem[Dunn(1961)]{dunn1961multiple}
Olive~Jean Dunn.
\newblock Multiple comparisons among means.
\newblock \emph{Journal of the American statistical association}, 56\penalty0
  (293):\penalty0 52--64, 1961.

\bibitem[Fraser(1968)]{fraser1968structural}
Donald~AS Fraser.
\newblock \emph{Structural inference}.
\newblock John Wiley \& Sons, New York, NY, 1968.

\bibitem[Hannig et~al.(2016)Hannig, Iyer, Lai, and Lee]{hannig2016generalized}
Jan Hannig, Hari Iyer, Randy~CS Lai, and Thomas~CM Lee.
\newblock Generalized fiducial inference: A review and new results.
\newblock \emph{Journal of the American Statistical Association}, 111\penalty0
  (515):\penalty0 1346--1361, 2016.

\bibitem[Houssineau(2018)]{houssineau2018parameter}
Jeremie Houssineau.
\newblock Parameter estimation with a class of outer probability measures.
\newblock \emph{arXiv preprint arXiv:1801.00569}, 2018.

\bibitem[Kong(1987)]{kong1987multivariate}
Chung Tung~Augustine Kong.
\newblock \emph{Multivariate Belief Functions and Graphical Models}.
\newblock PhD thesis, Harvard University, 1987.

\bibitem[Martin and Liu(2015)]{martin2015inferential}
Ryan Martin and Chuanhai Liu.
\newblock \emph{{Inferential Models}: reasoning with uncertainty}.
\newblock CRC Press, Boca Raton, FL, 2015.

\bibitem[Shafer(1976)]{shafer1976mathematical}
Glenn Shafer.
\newblock \emph{A mathematical theory of evidence}.
\newblock Princeton University Press, Princeton, NJ, 1976.

\end{thebibliography}
% should be alandvonvereweg19.bib if this were a submission

\end{document}